\documentclass[11pt]{article}
\usepackage[left=1in,top=1in,right=1in,bottom=1in,head=.1in,nofoot]{geometry}

\usepackage{setspace,url,bm,amsmath} 
\usepackage{subcaption}

\usepackage{titlesec} 
\titlelabel{\thetitle.\quad} 
\titleformat*{\section}{\bf\large\center} 

\usepackage{graphicx} 
\usepackage{bbm}
\usepackage[export]{adjustbox}
\usepackage{latexsym}
\usepackage{mathrsfs}
\usepackage{amssymb}
\usepackage{authblk}
\usepackage{hyperref}
\usepackage[table]{xcolor}

\usepackage{tikz}
\usepackage{enumitem}

\usepackage{amsthm}
\theoremstyle{definition}

\newtheorem*{theorem*}{Theorem}
\newtheorem{theorem}{Theorem}

\newtheorem{lemma}{Lemma}
\newtheorem{remark}{Remark}

\newtheorem*{note*}{Note}

\newtheorem{corollary}{Corollary}
\newtheorem*{corollary*}{Corollary}

\usepackage{natbib} 
\bibpunct{(}{)}{;}{a}{}{,} 

\usepackage{etoolbox} 
\apptocmd{\sloppy}{\hbadness 10000\relax}{}{} 

\usepackage{color}
\usepackage{listings}

\newcommand{\ind}{\perp \! \! \! \perp}

\begin{document}
\doublespacing
\title{\bf Sharp bounds on the relative treatment effect for ordinal outcomes}

\author{Jiannan Lu, Yunshu Zhang and Peng Ding\thanks{Jiannan Lu is Senior Data Scientist (E-mail: jiannl@microsoft.com), Analysis and Experimentation, Microsoft Corporation, Redmond, WA 98052, U.S.A. Yunshu Zhang is Doctoral Student (E-mail: yzhan234@ncsu.edu), Department of Statistics, North Carolina State University, Raleigh, NC 27695, U.S.A. Peng Ding is Assistant Professor (E-mail: pengdingpku@berkeley.edu), Department of Statistics, University California, Berkeley, CA 94270, U.S.A. }}

\date{}
\maketitle
\begin{abstract}
For ordinal outcomes, the average treatment effect is often ill-defined and hard to interpret. Echoing \citet{Agresti:2017}, we argue that the relative treatment effect can be a useful measure especially for ordinal outcomes, which is defined as $\gamma = \mathrm{pr}\{  Y_i(1) > Y_i(0) \} - \mathrm{pr}\{ Y_i(1) < Y_i(0)  \}$, with $Y_i(1)$ and $Y_i(0)$ being the potential outcomes of unit $i$ under treatment and control, respectively. Given the marginal distributions of the potential outcomes, we derive the sharp bounds on $\gamma,$ which are identifiable parameters based on the observed data. \citet{Agresti:2017} focused on modeling strategies under the assumption of independent potential outcomes, but we allow for arbitrary dependence.  
\end{abstract}


\section{Causal inference with ordinal outcomes}\label{sec::intro}

Ordinal outcomes are very common in empirical research \citep[e.g.,][]{Whitehead:2001, Scharfstein:2004, Huang:2017, Liu:2018}.
Consider a binary treatment and an ordinal outcome with labels $0,\ldots,J-1$, where 0 and $J-1$ denote the worst and best categories, respectively. Define $\left\{Y_i(1), Y_i(0)\right\}$ as the potential outcomes of  unit $i \in \{1, \ldots, N\}$ under treatment and control, respectively.
For all
$
k,l=0, \ldots, J-1,
$
let 
$
p_{kl} = \mathrm{pr} \left\{ Y_i(1) = k, Y_i(0) = l \right\} 
$
denote the probability that the potential outcome is $k$ under treatment and $l$ under control, respectively. 
The probability matrix
$
\bm P = (p_{kl})_{0 \le k, l \le J-1}
$
characterizes the joint distribution of the potential outcomes. Let
$
p_{k+} = \sum_{l^\prime=0}^{J-1}p_{kl^\prime}
$
and
$
p_{+l} = \sum_{k^\prime =0}^{J-1}p_{k^\prime l}
$
be the marginal distributions of the potential outcomes under treatment and control, respectively. We let $\bm p_1 = ( p_{0+}, \ldots, p_{J-1, +} )^\textrm{T}$ and $\bm p_0 = ( p_{+0}, \ldots, p_{+, J-1} )^\textrm{T}$ denote the marginal probability vectors.

For ordinal outcomes, the average treatment effect $E\{ Y_i(1) - Y_i(0)  \}$ is often hard to interpret, if there is no clear definition of ``distance'' between different categories. In contrast, the parameters $\tau = \mathrm{pr} \{ Y_i(1) \geq  Y_i(0) \} $ and $\eta  = \mathrm{pr} \{ Y_i(1) > Y_i(0) \} $ have clear interpretations as the probabilities that the treatment is beneficial and strictly beneficial for the outcome \citep{Newcombe:2006a, Newcombe:2006b, Zhou:2008, Huang:2017, Lu:2018}. Recently, \citet{Agresti:2017} used the relative treatment effect for ordinal outcomes \citep{Agresti:2010}, defined as
\begin{align}
\label{eq:alpha}
\gamma 
&= \mathrm{pr} \{ Y_i(1) > Y_i(0) \} - \mathrm{pr} \{ Y_i(1) < Y_i(0) \}   \nonumber  \\
&= \mathop{\sum\sum}_{ k > l}p_{kl}  - \mathop{\sum\sum}_{ k < l}p_{kl}.
\end{align}
We can verify that 
$
\gamma 
= \mathrm{pr} \{ Y_i(1) > Y_i(0) \} - [ 1 - \mathrm{pr} \{ Y_i(1) \geq  Y_i(0) \}]
= \tau  +\eta - 1 .
$
The parameters $\tau$, $\eta$ and $\gamma$ are closely related to the classic Wilcoxon--Mann--Whitney statistic for testing equality of two distributions \citep{Kruskal:1952, Kruskal:1957, Klotz:1966, Vargha:1998, Chung:2016, Divine:2018}. The parameters $\tau$, $\eta$ and $\gamma$ depend on the joint distribution of the potential outcomes and are not identifiable based on the observed data \citep{Hand:1992, Demidenko:2016, Huang:2017, Lu:2018, Greenland:2019}.
\citet{Huang:2017} obtained numerical bounds on $\tau$ and $\eta$, and \citet{Lu:2018} derived explicit formulas of these bounds. \citet{Agresti:2017} and \citet{Cheng:2009} discussed $\gamma$ assuming independent potential outcomes implicitly and explicitly. \citet{Chiba:2018} proposed a Bayesian approach to infer $\gamma$, which requires imposing a prior on the joint distribution of the potential outcomes. \cite{Fay:2018} and \cite{Fay:2018b} pointed out the non-identifiability nature of $\gamma$ and proposed a non-sharp bound on $\gamma$ given the marginal distributions of $\mathrm{pr}\{ Y_i(1) \}$ and $\mathrm{pr}\{ Y_i(0) \},$ based on \cite{Lu:2018}'s bounds on $\tau$ and $\eta.$

For $J=2$ (i.e., when $Y$ is binary), the relative treatment effect reduces to
$
\gamma = p_{1+} - p_{+1} = E\{ Y_i(1) - Y_i(0)  \}  =  E\{ Y_i(1) \}  -  E\{ Y_i(0)  \},
$
which is actually the average treatment effect. Because the average treatment effect depends only on the marginal distributions of the potential outcomes, $\gamma$ is identifiable from the observed data with $J=2$. However, $\gamma$ becomes unidentifiable when $J \ge 3$, because it depends on the joint distribution of the treated and control potential outcomes. We adopt the partial identification strategy \citep[c.f.][]{Manski:2003, Richardson:2014} and focus on the sharp bounds on $\gamma$. We compute the maximum and minimum values of $\gamma$ that are compatible with the marginal distributions of the potential outcomes. As a theoretical starting point, we assume that the marginal probabilities $\bm p_1$ and $\bm p_0$ are known, and later we will incorporate sampling variability. The sharp upper bound $\gamma_U$ is the solution of the following linear programming problem:
\begin{equation*}
\begin{aligned}
\gamma_U
&= \underset{\bm P}{\max} && \mathop{\sum\sum}_{ k > l}p_{kl}  - \mathop{\sum\sum}_{ k < l}p_{kl} \\
& \quad\; \text{subject to}
&& \sum_{l^\prime = 0 }^{J-1} p_{k l^\prime} = p_{k +} \quad (k = 0, \ldots, J-1); \\
&&& \sum_{k^\prime = 0}^{J-1}p_{k^\prime l} = p_{+ l} \quad (l = 0, \ldots, J-1); \\
&&& p_{kl} \ge 0 \quad (k, l = 0,\ldots,J-1) . 
\end{aligned}
\end{equation*}
The sharp lower bound $\gamma_L$ is the corresponding minimum value subject to the same set of constraints. By definitions, the sharp upper and lower bounds are functions of the marginal probabilities $\bm p_1$ and $\bm p_0$, although the relative treatment effect $\gamma$ itself is a function of the joint probability matrix $\bm{P}.$ \citet{Balke:1997} and \citet{Huang:2017} used linear programming to obtain bounds on different causal parameters for ordinal and more general outcomes. Numerically, we can easily obtain the values of $\gamma_U$ and $\gamma_L$ for given values of $\bm p_1$ and $\bm p_0$. However, our goal here is to derive explicit formulas, as in \citet{Balke:1997} and \citet{Lu:2018}, which give more transparent interpretations and allow for convenient estimation and inference. 


The rest of the paper is organized as follows. Section \ref{sec:theory} derives the sharp bounds on the relative treatment effect. Section \ref{sec:inference} discusses the statistical inference based on the derived bounds under different scenarios such as completely randomized experiments and observational studies. Section \ref{sec:examples} presents two examples to illustrate our proposed method. We relegate all technical details to the supplementary material.

\section{Main results: Sharp bounds on $\gamma$}
\label{sec:theory}
 
\subsection{Notation} 
\label{subsec:notation}

We introduce a few quantities that are needed to express the sharp bounds on $\gamma$. For each fixed $j = 1, \ldots, J-1$ and $m=1, \ldots, J-j,$ let
\begin{equation}
\label{eq:delta_original}
\delta_{jm} 
= 
\sum_{k=j}^{J-1} p_{k+}
+  
\sum_{k=j+m}^{J-1} p_{k+} 
+
\sum_{l=0}^{j-2} p_{+l} 
-
\sum_{l=j+m-1}^{J-1} p_{+l}.
\end{equation}
Define the summation to be zero when the range is empty, e.g., $\sum_{l=0}^{j-2}p_{+l} =0$ if $j = 1.$ Importantly, the $\delta_{jm}$'s depend only on the marginal probabilities $\bm{p}_1$ and $\bm{p}_0.$ Before moving forward, we provide insights on the important roles the $\delta_{jm}$'s play in deriving the sharp bounds on the relative treatment effect $\gamma.$ For example, by taking the difference between
\begin{align*}
\mathrm{pr} \{ Y_i(1) > Y_i(0) \}
&= \mathrm{pr} \{ Y_i(1) > Y_i(0), Y_i(0) = 0 \} + \mathrm{pr} \{Y_i(1) > Y_i(0), Y_i(0) \ge 1 \} \\
&\le \mathrm{pr} \{ Y_i(1) \ge 1, Y_i(0) = 0 \} + \mathrm{pr} \{Y_i(1) \geq 2, Y_i(0) \ge 1 \} \\
&\le \mathrm{pr} \{ Y_i(1) \ge 1, Y_i(0) = 0 \} + \mathrm{pr} \{ Y_i(1) \ge 2 \} \\
&= \mathrm{pr} \{ Y_i(1) \ge 1 \} - \mathrm{pr} \{ Y_i(1) \ge 1, Y_i(0) \ge 1 \} + \mathrm{pr} \{ Y_i(1) \ge 2 \}
\end{align*}
and
\begin{align*}
\mathrm{pr} \{ Y_i(1) < Y_i(0) \}
& \ge \mathrm{pr} \{ Y_i(1) < Y_i(0), Y_i(1) = 0 \} \\
&\ge 
\mathrm{pr} \{ Y_i(0) \ge 1, Y_i(1) = 0 \} \\
&= \mathrm{pr} \{ Y_i(0) \ge 1 \} - \mathrm{pr} \{ Y_i(1) \ge 1, Y_i(0) \ge 1 \},
\end{align*}
we obtain   
\begin{align*}
\gamma 
&\le
\mathrm{pr} \{ Y_i(1) \ge 1\}
+ 
\mathrm{pr} \{ Y_i(1) \ge 2 \}
- 
\mathrm{pr} \{ Y_i(0) \ge 1\}
\\
&= 
\sum_{k=1}^{J-1} p_{k+} 
+ 
\sum_{k=2}^{J-1} p_{k+}
-
\sum_{l=1}^{J-1} p_{+l} \\
&= 
\delta_{11}.
\end{align*}
In other words, $\delta_{11}$ is a loose upper bound on $\gamma.$ Similarly, we can prove that other $\delta_{jm}$'s are also loose upper bounds on $\gamma.$ Interestingly, in the next subsection we will show that the $\delta_{jm}$'s together can sharply bound $\gamma$.

\subsection{Main theorem, corollaries and remarks}
\label{subsec:main-retults}

We now present the main result of this paper.

\begin{theorem} 
\label{thm:main}
When $J \ge 3$, the sharp upper bound on the relative treatment effect $\gamma$ is
\begin{equation}
\label{eq:alpha-bound}
\gamma _U = 
\min_{1 \le j \le J - 1}  ~ \min_{ 1 \le m \le J - j} \delta_{jm}. 
\end{equation}
\end{theorem}

In the supplementary material we provide a proof of Theorem \ref{thm:main}, which consists of two parts. First, as previously mentioned, we show that $\gamma _U \leq \delta_{jm}$ for $j = 1, \ldots, J-1$ and $m=1, \ldots, J-j.$ Second, we prove the sharpness of $\gamma_U$ by directly constructing a probability matrix $\bm{P}$ attaining the bound given the marginal distributions. Although not affecting the proof, it is worth noting that the probability matrix attaining $\gamma_U$ might not be unique in general.

By switching the labels of the treatment and control potential outcomes, it is straightforward to obtain the sharp lower bound on the relative treatment effect $\gamma.$ 

\begin{corollary} 
\label{coro:sharp-lower-bound}
When $J \ge 3$, the sharp lower bound on the relative treatment effect $\gamma$ is 
\begin{equation}
\label{eq:alpha-lower-bound}
\gamma_L = \max_{1 \le j \le J - 1}~ \max_{ 1 \le m \le J - j} \xi_{jm},
\end{equation}
where for all $j=1, \ldots, J-1$ and $m = 1, \ldots, J-j,$
\begin{equation}
\label{eq:xi_original}
\xi_{jm} 
= 
\sum_{k=j+m-1}^{J-1} p_{k+}
-
\sum_{l=j}^{J-1} p_{+l}
-  
\sum_{l=j+m}^{J-1} p_{+l} 
-
\sum_{k=0}^{j-2} p_{k+} 
,
\end{equation}
with summations being zero if the range is empty. 
\end{corollary}

\begin{remark}
For $J = 3$, we can verify that 
$
\delta_{11} = p_{1+} + 2p_{2+} - p_{+1} - p_{+2},
$ 
$
\delta_{12} = p_{2+} - p_{+2} + p_{+1},
$
and 
$
\delta_{21} = p_{2+} - p_{+2} + p_{+0},
$ 
and that 
$
\xi_{11} = p_{1+} + p_{2+} - p_{+1} - 2p_{+2}, 
$
$
\xi_{12} = p_{2+} - p_{+1} - p_{+2},
$
and 
$
\xi_{21} = p_{2+} - p_{+2} - p_{0+}. 
$
Consequently, the sharp lower bound in Theorem \ref{thm:main} reduces to
$
\gamma_L 
= 
\max
\left( 
p_{1+} + p_{2+} - p_{+1} - 2p_{+2}
,    
p_{2+} - p_{+1} - p_{+2}
,      
p_{2+} - p_{+2} - p_{0+}
\right),
$
and the sharp upper bound in Corollary \ref{coro:sharp-lower-bound} reduces to
$
\gamma_U
= 
\min
\left( 
p_{1+} + 2p_{2+} - p_{+1} - p_{+2}
,   
p_{2+} - p_{+2} + p_{+1}
,   
p_{2+} - p_{+2} + p_{+0}
\right).
$
\end{remark}

Intuitively, $\gamma_U$ and $\gamma_L$ correspond to ``extremely'' positive and negative associations between potential outcomes $Y_i(1)$ and $Y_i(0).$  In practice, because they are characteristics of the same unit, it is plausible to rule out the scenarios with negatively associated potential outcomes \citep{Ding:2016, Lu:2018}. Therefore, we can use the previous result with independent potential outcomes as a lower bound \citep{Cheng:2009, Agresti:2010, Agresti:2017}. 

\begin{corollary}
With independent potential outcomes, i.e., $p_{kl} = p_{k+}p_{+l}$, the relative treatment effect can be identified as
$
\gamma_I  =  \mathop{\sum\sum}_{ k > l}p_{k+}p_{+l}  - \mathop{\sum\sum}_{ k < l}p_{k+}p_{+l}.
$
\end{corollary}

We suggest using $ [  \gamma_I , \gamma_U ]$ as the bounds on $\gamma$ as in the examples in Section \ref{sec:examples}.

\section{Statistical modeling and inference} 
\label{sec:inference}

\subsection{Point estimation}
\label{subsec:point-estimate}

To estimate the sharp bounds of the relative treatment effect $\gamma,$ we first estimate the marginal probabilities of the potential outcomes. Let $Z_i$ be the binary treatment indicator, with $Z_i=1$ if unit $i$ receives treatment and $Z_i=0$ if unit $i$ receives control. The observed outcome is therefore $Y_i = Z_i Y_i(1) + (1-Z_i) Y_i(0)$. In some studies, we also have pretreatment covariates $\bm X_i$. We assume that the observations $\{  Z_i, \bm X_i, Y_i(1), Y_i(0) \}_{i=1}^N$ are independent and identically draws from a super population. Following \cite{Lu:2018}, we consider the following two scenarios:

\begin{enumerate}

\item Completely randomized experiment with $Z_i\ind \{ Y_i(1), Y_i(0)  \}$. Therefore, we can estimate the marginal probabilities by their sample analogues
$$
\widehat{p}_{k+} = \frac{\sum_{i=1}^N Z_i \bm{1}_{\{Y_i = k\}}}{\sum_{i=1}^N Z_i},
\quad
\widehat{p}_{+l} = \frac{\sum_{i=1}^N (1 - Z_i) \bm{1}_{\{Y_i = l\}}}{\sum_{i=1}^N (1 - Z_i)} . 
$$
    
\item Unconfounded observational study with $Z_i\ind \{ Y_i(1), Y_i(0)  \}\mid \bm X_i$. For illustration, we focus on the propensity score weighting and outcome modeling approaches. First, we can estimate the marginal probabilities by the inverse propensity score weighting:
$$
\widehat p_{k+}  = \sum_{i=1}^N \frac{Z_i  \bm{1}_{\{Y_i=k\}} }{  \widehat{e}(\bm X_i)  } \big / \sum_{i=1}^N \frac{Z_i   }{  \widehat{e}(\bm X_i)  },
\quad
\widehat{p}_{+l}  = \sum_{i=1}^N \frac{(1-Z_i)  \bm{1}_{\{Y_i=l\}} }{ 1- \widehat{e}(\bm X_i)  } \big / \sum_{i=1}^N \frac{1-Z_i   }{ 1- \widehat{e}(\bm X_i)  },
$$
where $\widehat{e}(\bm X_i)$ is the fitted value of the propensity score $e(\bm X_i) = \mathrm{pr}(Z_i=1\mid \bm X_i)$, for example, via a logistic regression of the treatment indicator on the covariates. Second, we can fit two outcome models $ \mathrm{pr}(Y_i\mid Z_i=1, \bm X_i)$  and $ \mathrm{pr}(Y_i\mid Z_i=0, \bm X_i)$ using the data under treatment and control, respectively. A canonical choice for ordinal outcomes is the proportional odds model \citep[c.f.][]{Agresti:2010}. We then obtain the fitted values 
$
\widehat{p}_{k+}(\bm X_i) =  \widehat{\mathrm{pr}}(Y_i=k\mid Z_i=1, \bm X_i)
$
and
$
\widehat{p}_{+l}(\bm X_i) =  \widehat{\mathrm{pr}}(Y_i=l\mid Z_i=0, \bm X_i)
$ 
for all units. The final outcome-regression estimators for the marginal probabilities are
$
\widehat{p}_{k+}  = \sum_{i=1}^N\widehat{p}_{k+}(\bm X_i) / N
$
and
$
\widehat{p}_{+l}  = \sum_{i=1}^N\widehat{p}_{+l}(\bm X_i) / N. 
$
We can estimate the bounds $[\gamma_I, \gamma_U]$ using a plug-in approach after obtaining the $\widehat{p}_{k+}$'s and $\widehat{p}_{+l} $'s.

\end{enumerate}

\subsection{Sharpening bounds using covariates}
\label{subsec:sharpening}

\citet{Agresti:2017}'s strategy of covariate adjustment is slightly different from the above discussion in Section \ref{subsec:point-estimate}. \citet{Agresti:2017} first estimated the conditional relative treatment effect given covariates, and then averaged over the empirical distribution of covariates. This is similar to the strategy of using covariates to sharpen the bounds \citep{Grilli:2008, Lee:2009, Long:2013, Lu:2018}. In particular, we can first estimate the conditional bounds given covariates
$
\widehat{\gamma}_I(\bm X_i) =   \mathop{\sum\sum}_{ k > l} \widehat{p}_{k+}(\bm X_i) \widehat{p}_{+l}(\bm X_i) 
 - \mathop{\sum\sum}_{ k < l} \widehat{p}_{k+}(\bm X_i) \widehat{p}_{+l}(\bm X_i)
$
and
$
\widehat{\gamma}_U(\bm X_i) = \min_{1 \le j \le J - 1}  ~ \min_{ 1 \le m \le J - j} \widehat{\delta}_{jm}(\bm X_i),
$
and then estimate the bounds by 
$
\widehat{\gamma}_I  = \sum_{i=1}^N \widehat{\gamma}_I(\bm X_i) / N
$ 
and 
$
\widehat{\gamma}_U  = \sum_{i=1}^N \widehat{\gamma}_U(\bm X_i) / N.
$

\subsection{Confidence intervals}


Following the existing literature on statistical inferences for partially identified parameters \citep{Cheng:2006, Yang:2016}, we construct a $(1-\alpha)$-level confidence interval for the sharp bounds $(\gamma_I, \gamma_U),$ which automatically covers $\gamma$ at least $100(1-\alpha)\%$ of the time. However, as pointed out by \cite{Hirano:2012}, delicate issues arise in this case, especially the trade-off between simplicity of implementation and uniformity of the coverage properties of confidence intervals. For the empirical examples in Section \ref{sec:examples}, we employ \cite{Horowitz:2000}'s non-parametric bootstrap interval
$
\{
\widehat \gamma_I - z^*_\alpha
,
\widehat \gamma_U + z^*_\alpha
\},
$
where we obtain the threshold $z^*_\alpha$ by solving the equation
$
\mathrm{pr}_B
\{
\widehat \gamma_I^* - z^*_\alpha
\le
\widehat \gamma_I, \widehat \gamma_U
\le
\widehat \gamma_U^* + z^*_\alpha
\}
=
1 - \alpha,
$
where $\widehat \gamma_I^*$ and $\widehat \gamma_U^*$ are drawn from the Bootstrap distribution $\mathrm{pr}_B.$ While more sophisticated methods \citep[e.g.,][]{Romano:2010, Chernozhukov:2013, Jiang:2018} may be more rigorous theoretically, previous discussions \citep[e.g.,][]{Lu:2018} showed that the interval by \cite{Horowitz:2000} achieved similar finite-sample performances, at least in the context of ordinal outcomes.

\section{Applications}
\label{sec:examples}

\subsection{A randomized experiment}

We illustrate our theory and method using the Sexual Assault Resistance Education Trial \citep{Senn:2015}, previously analyzed by \cite{Lu:2018}. In this randomized experiment, the treatment is the enhanced Assess, Acknowledge and Act program, which aims at preventing sexual assaults. The outcome of interest has six categories from ``complete rape'' to ``no reporting of any non-consensual sexual contact,'' labelled as 0--5.
The numbers of units are $(23 , 15 , 48  , 67 , 121 , 177 , 451)$ in the treatment arm and $(42 , 40 , 62  , 103  , 184  , 11 ,  442)$ in the control arm, corresponding to the outcome categories $(0,1,2,3,4,5)$. 
Based on these data, we estimate the sharp bounds on $\gamma$ as 
$
[ \widehat{\gamma}_I, \widehat{\gamma}_U] = [0.387, 0.900],
$ 
and the corresponding 95\% bootstrap confidence interval is 
$
[0.315, 0.972].
$
The results imply that the program is beneficial, which corroborate the recommendations by \cite{Senn:2015} and \cite{Lu:2018}.


\subsection{An observational study}

We illustrate our theory and method using an observational study from the Karolinska Institute in Stockholm, Sweden, which was previously analyzed by \citet{Rubin:2008}. The data have 158 cardia cancer patients diagnosed between 1988 and 1995. The treatment is whether the patient is diagnosed in a high volume hospital, defined as treating more than 10 patients with cardia cancer during that period. The outcome is the survival time of the patient after the diagnosis, with three categories ordered as ``one year,'' ``between two and four years'' and ``longer than five years''. For patients diagnosed in a high volume hospital, 51 survived for one year, 18 survived between two and four years, and 10 survived longer than five year. For patients diagnosed in a low volume hospital, the numbers are 50, 21 and 8.  Pre-treatment covariates include the age at diagnosis, indicator of male, and indicator of whether the patient is from the rural areas. 
The last covariate is an important confounder in this example, because patients from rural areas would be more likely to attend low volume hospitals ($p$-value 0.0001).

We assume that the treatment is unconfounded given the observed pre-treatment covariates. We first fit two separate proportional odds models for the outcomes under treatment and control, respectively. We then obtain the fitted probabilities for each individual under both treatment and control. We finally use the strategy in Section \ref{subsec:sharpening} to obtain sharp bounds on the relative treatment effect as 
$
[ \widehat{\gamma}_I, \widehat{\gamma}_U] = [0.055, 0.183]
$ 
with the 95\% bootstrap confidence interval 
$
[-0.137, 0.375]. 
$
The lower confidence limit, corresponding to independent potential outcomes as in \citet{Agresti:2017}, is smaller than 0 although the point estimate of the lower bound is positive. 

%
%

\section*{Acknowledgements}

The authors thank the co-Editor and a reviewer for their constructive comments. 
This work is motivated by an open question in Lu's doctoral thesis, and he gratefully acknowledges his advisors, Professors Tirthankar Dasgupta, Joseph Blitzstein and Luke Miratrix. Zhang thanks Professor Ke Deng  for  valuable suggestions. 
Ding is partially supported by Institute of Education Sciences Grant R305D150040 and National Science Foundation Grant DMS-1713152.

\bibliographystyle{apalike}
\bibliography{rte.bib}

\section*{Supporting Information}
Web Appendices referenced in Section \ref{sec::intro}, R code, and data are available with this paper at the Biometrics website on Wiley Online Library.

\newpage

\setcounter{equation}{0}
\setcounter{section}{0}
\setcounter{page}{1}

\renewcommand {\theequation} {S\arabic{equation}}
\renewcommand {\thesection} {S\arabic{section}}

\begin{center}
{\Large \bfseries 
Supporting Information for ``Sharp bounds on the relative treatment effect for ordinal outcomes''}\\
\bigskip
by Jiannan Lu, Yunshu Zhang and Peng Ding
\end{center}

\section{Overview and notation}

The supplementary materials are organized in the following way. Section \ref{sec:lemmas} gives several lemmas that are useful for proving the main results. Section \ref{sec:theorem} gives a proof of Theorem \ref{thm:main}, and Section \ref{sec:corollary} gives a proof of Corollary \ref{coro:sharp-lower-bound}. 

To simplify the proofs, we need the distributional causal effects
\begin{equation}
\label{eq:Delta}
\Delta_j = \mathrm{pr}\left\{ Y_i(1) \ge j \right\} - \mathrm{pr}\left\{ Y_i(0) \ge j \right\}
= \sum_{k \ge j} p_{k+}  -  \sum_{l \ge j} p_{+l},\quad (j=1, \ldots, J-1),
\end{equation}
which compare the marginal distribution functions of the potential outcomes. By \eqref{eq:delta_original}, \eqref{eq:xi_original} and \eqref{eq:Delta}, for all $j = 1, \ldots, J-1$ and $m = 1, \ldots, J-j,$
\begin{equation}
\label{eq:delta}
\delta_{jm} 
= \Delta_j 
+   \sum_{l=0}^{j-2}p_{+l} 
+  \sum_{k=j+m}^{J-1} p_{k+} 
+   \sum_{l=j}^{j+m-2} p_{+l}
\end{equation}
and
\begin{equation}
\label{eq:xi}
\xi_{jm} 
= \Delta_j 
-   \sum_{k=0}^{j-2}p_{k+} 
-   \sum_{l=j+m}^{J-1} p_{+l} 
-   \sum_{k=j}^{j+m-2} p_{k+},
\end{equation}
Again, we follow the convention in the main text to define the summation as zero when the range is empty, e.g., $\sum_{l=j}^{j+m-2}p_{+l} =0$ if $m = 1.$

\section{Lemmas and their proofs}\label{sec:lemmas}

In this section, we introduce three lemmas, which play instrumental roles in deriving the sharp bounds on the relative treatment effect $\gamma.$ 

\subsection{Lemma \ref{le:1} from \cite{Lu:2018}}


\begin{lemma}
\label{le:1}
Assume that $\left( x_0, \ldots, x_{n-1} \right)$ and $\left( y_0, \ldots, y_{n-1} \right)$ are non-negative constants.
\begin{enumerate}[label= (\alph*), ref = \ref{le:1}(\alph*)]
\item \label{le:1-a}
If
$
\sum_{r=s}^{n-1} x_r \ge \sum_{r=s}^{n-1} y_r
$
for all $s = 0, \ldots, n-1,$ there exists an $n\times n$ lower triangular matrix
$
\bm A_n = (a_{kl})_{0 \le k,l \le n-1}
$
with non-negative elements such that
\begin{equation*}
\sum_{l^\prime=0}^{n-1} a_{kl^\prime} \le x_k,
\quad
\sum_{k^\prime=0}^{n-1} a_{k^\prime l}= y_l
\quad
(k, l = 0, \ldots, n-1).
\end{equation*}

\item \label{le:1-c}
If
$
\sum_{r=0}^s x_r \le \sum_{r=0}^s y_r
$
for all $s = 0, \ldots, n-1,$
there exists an $n\times n$ lower triangular matrix
$
\bm B_n = (b_{kl})_{0 \le k,l \le n-1}
$
with non-negative elements such that
\begin{equation*}
\sum_{l^\prime=0}^{n-1} b_{k l^\prime}= x_k,
\quad
\sum_{k^\prime=0}^{n-1} b_{k^\prime l} \le y_l
\quad
(k, l = 0, \ldots, n-1).
\end{equation*}
\end{enumerate}
\end{lemma}


\subsection{Lemma \ref{le:2} and its proof}

The second lemma establishes various relationships among the $\delta_{jm}$'s defined in \eqref{eq:delta}.

\begin{lemma}
\label{le:2}
For fixed $j = 1, \ldots, J-2,$
\begin{enumerate}[label= (\alph*), ref = \ref{le:2}(\alph*)]
\item \label{le:2-a}
$
\delta_{jm} + p_{ + ,j + m - 1} - p_{j + m, + } = \delta_{j,m + 1}
$
for $m = 1, \ldots, J - 1 - j ; $

\item \label{le:2-b}
$
\delta_{j + 1,m} + p_{j + } - p_{ + ,j - 1} = \delta_{j,m + 1}
$
for $m = 1, \ldots, J - 1 - j ;$

\item \label{le:2-c}
$
\delta_{j + 1,J - 1 - j} + p_{ + ,J - 1} - p_{0 + } = \delta _{1j}.
$

\end{enumerate}
\end{lemma}

\medskip
\begin{proof}[Proof of Lemma \ref{le:2-a}]
Notice that
$$
\sum_{l=j}^{j+m-1} p_{+l}
= \sum_{l=j}^{j+m-2} p_{+l} + p_{+, j+m-1},
\quad
\sum_{k=j+m+1}^{J-1} p_{k+}
= \sum_{k=j+m}^{J-1} p_{k+} - p_{j+m, +}.
$$
Therefore,
\begin{align*}
\delta_{j,m + 1}
&= \Delta_j + \sum_{l=0}^{j-2}p_{+l} 
+ \sum_{k=j+m+1}^{J-1} p_{k+} 
+  \sum_{l=j}^{j+m-1} p_{+l} \\
&= \Delta_j + \sum_{l=0}^{j-2}p_{+l} 
+  \sum_{k=j+m}^{J-1} p_{k+} - p_{j+m, +}
+ \sum_{l=j}^{j+m-2} p_{+l} + p_{+, j+m-1}\\
&= \delta_{jm} + p_{+, j+m-1} - p_{j+m, +}.
\end{align*}
The proof is complete.
\end{proof}

\smallskip
\begin{proof}[Proof of Lemma \ref{le:2-b}]
Notice that
$$
\sum_{l=0}^{j-2}p_{+l} = \sum_{l=0}^{j-1}p_{+l} - p_{+, j-1},
\quad 
\sum_{l=j}^{j+m-1} p_{+l} = \sum_{l=j+1}^{j+m-1} p_{+l} + p_{+j},
$$
and that
$
{\Delta _j} = {\Delta _{j + 1}} + {p_{j + }} - {p_{ + j}}
$
by \eqref{eq:Delta}. Therefore,
\begin{align*}
\delta_{j,m + 1}
&= \Delta_j + \sum_{l=0}^{j-2}p_{+l} 
+ \sum_{k=j+m+1}^{J-1} p_{k+} + \sum_{l=j}^{j+m-1} p_{+l} \\
&= \Delta_j + \sum_{l=0}^{j-1}p_{+l} - p_{+, j-1}
+ \sum_{k=j+m+1}^{J-1} p_{k+} + \sum_{l=j+1}^{j+m-1} p_{+l} + p_{+j} \\
&= \Delta_{j+1} + p_{j+} + \sum_{l=0}^{j-1}p_{+l} - p_{+, j-1} + \sum_{k=j+m+1}^{J-1} p_{k+} + \sum_{l=j+1}^{j+m-1} p_{+l} \\
&= \delta_{j+1, m} + p_{j+} - p_{+, j-1}.
\end{align*}
The proof is complete.
\end{proof}

\smallskip
\begin{proof}[Proof of Lemma \ref{le:2-c}]
By repeatedly utilizing Lemma \ref{le:2-b}, we have 
\begin{align*}
\delta_{j+1, J-1-j} 
= \delta_{j, J-j} + (p_{+, j-1} - p_{j+}) 
= \cdots 
= 
\delta_{1, J-1} + \sum_{l=0}^{j-1} p_{+l} - \sum_{k=1}^j p_{k+}.
\end{align*}
Moreover, by repeatedly utilizing Lemma \ref{le:2-a}, we have 
\begin{align*}
\delta_{1, J-1}
= \delta_{1, J-2} + (p_{+, J-2} - p_{J-1, +}) 
= \cdots 
= \delta_{1j} + \sum_{l=j}^{J-2}p_{+l} - \sum_{k=j+1}^{J-1}p_{k+}.
\end{align*}
Combining the above two equations, we have 
\begin{align*}
\delta_{j+1, J-1-j} 
= \delta_{1j} + (1-p_{+, J-1}) - (1-p_{0+}) 
= \delta_{1j} + p_{0+} - p_{+, J-1},
\end{align*}
which completes the proof.
\end{proof}

\subsection{Lemma \ref{le:3} and its proof}

Lemma \ref{le:3}  bridges the first two lemmas by repeatedly utilizing Lemma \ref{le:2} to find subsets of the marginal probabilities which meet the conditions of Lemma \ref{le:1}. When proving the main theorem, we utilize Lemma \ref{le:3}  to construct a probability matrix attaining the upper bound $\gamma_U.$

\begin{lemma}
\label{le:3}
Let
$
\Omega = \left\{ (1, 1), \ldots, (1, J-1); (2, 1), \ldots, (2, J-2); \ldots; (J-1, 1) \right\}
$
denote the lexicographically ordered set of the 2-tuples $(j, m)$'s, where for each $j = 1, \ldots, J-1,$ the corresponding $m$ takes values between $1$ and $J-j.$ Let
\begin{equation}
\label{eq:the-minimum-tuple}
(j_1, m_1) = 
\min
\left\{ 
(j^\prime, m^\prime) \in \Omega:  
\delta_{j^\prime m^\prime} 
=  
\min_{1 \le j \le J - 1}~ \min_{ 1 \le m \le J - j} \delta_{jm}  
\right\}
\end{equation}
be the first 2-tuple attaining the minimum value of $\delta_{jm}$, and
\begin{equation}
\label{eq:define-lambda}    
\lambda_1 = \sum_{k=j_1}^{j_1+m_1-1} p_{k+} - \sum_{l=j_1-1}^{j_1+m_1-2} p_{+l}.
\end{equation} 
The following results hold. 
\begin{enumerate}[label= (\alph*), ref = \ref{le:3}(\alph*)]

\item \label{le:3-a}
If $j_1 > 1,$ let $\Omega_1 = \{1, \ldots, j_1-1\}$ and
\begin{equation}
\label{eq:define-q-1}
q_{k+} = p_{k+}
\quad
(k \in \Omega_1 \backslash \{j_1-1\} );
\quad
q_{j_1-1, +} = p_{j_1-1, +} + \min(0, \lambda_1). 
\end{equation}
Then
\begin{equation}    
\label{eq:constrtaints-marginal-final-1}
\sum_{l = n-1}^{j_1-2} p_{+l} \le \sum_{k = n}^{j_1-1} q_{k+}
\quad
(n=1, \ldots, j_1-1). 
\end{equation}

\item \label{le:3-b}
If $m_1 > 1,$ let $\Omega_2 = \{j_1+1, \ldots, j_1+m_1-1\}$ and
\begin{equation}
\label{eq:define-q-2}
q_{k+} = p_{k+}
\quad
(k \in \Omega_2\backslash\{j_1+m_1-1\});
\quad
q_{j_1+m_1-1, +} = p_{j_1+m_1-1, +} - \max(0, \lambda_1). 
\end{equation}
Then
\begin{equation}    
\label{eq:constrtaints-marginal-final-2}
\sum_{l=j_1+n-1}^{j_1+m_1-2} p_{+l} \le \sum_{k=j_1+n}^{j_1+m_1-1} q_{k+}
\quad
(n = 1, \ldots, m_1-1). 
\end{equation}

\item \label{le:3-c}
If $j_1 + m_1 < J,$ let $\Omega_3 = \{j_1+m_1-1, \ldots, J-2\}$ and 
\begin{equation}
\label{eq:define-q-3}
q_{+, j_1+m_1-1} = p_{+, j_1+m_1-1} - \max(0, \lambda_1); 
\quad
q_{+l} = p_{+l}
\quad
(l \in \Omega_3\backslash\{j_1+m_1-1\}).
\end{equation}
Then
\begin{equation}    
\label{eq:constrtaints-marginal-final-3}
\sum_{k=j_1+m_1}^{j_1+n-1} p_{k+} \le \sum_{l=j_1+m_1-1}^{j_1+n-2} q_{+l}
\quad
(n = m_1+1, \ldots, J-j_1).
\end{equation}

\end{enumerate}
\end{lemma}

\medskip
\begin{proof}[Proof of Lemma \ref{le:3-a}]
The starting point of the proof is that $\delta_{j_1, m_1}$ is the smallest among all the $\delta_{jm}$'s. Then, the key idea is to use Lemma \ref{le:2} to transform  
$
\{ 
\delta_{j_1, m_1} \le \delta_{j, m} : 
j = 1, \ldots, J-1; 
m = 1, \ldots, J-j
\}
$
into inequalities regarding certain subsets of the marginal probabilities. To be specific, if $j_1 > 1,$ we repeatedly utilize Lemma \ref{le:2-b} and obtain
\begin{equation}
\label{eq:delta-iteration-1}
\delta_{j_1, m_1}
= \delta_{n, j_1+m_1-n} + \sum_{s = n}^{j_1-1} (p_{+, s-1} - p_{s+})
\quad
(n=1, \ldots, j_1-1).
\end{equation}
By \eqref{eq:the-minimum-tuple},
$
\delta_{j_1, m_1}
\le
\delta_{n, j_1+m_1-n}
$
for
$
n = 1, \ldots, j_1 - 1,
$
implying
\begin{equation}
\label{eq:constrtaints-marginal-2}
\sum_{l = n-1}^{j_1-2} p_{+l} \le \sum_{k = n}^{j_1-1} p_{k+}
\quad
(n=1, \ldots, j_1-1).
\end{equation}
Moreover, by repeatedly utilizing Lemma \ref{le:2-a}, we have 
\begin{equation}
\label{eq:delta-iteration-1-2}
\delta_{n, j_1+m_1-n} 
= \delta_{n, j_1-n} - \lambda_1
\quad
(n=1, \ldots, j_1-1).
\end{equation}
By combining \eqref{eq:delta-iteration-1} and \eqref{eq:delta-iteration-1-2}, we have 
\begin{equation*}
\delta_{j_1, m_1}
= \delta_{n, j_1-n} + \sum_{s = n}^{j_1-1} (p_{+, s-1} - p_{s+}) - \lambda_1
\quad
(n=1, \ldots, j_1-1).
\end{equation*}
Similarly, because
$
\delta_{j_1, m_1} \le \delta_{n, j_1-n}
$
for all
$
n = 1, \ldots, j_1 - 1,
$
\begin{equation}    
\label{eq:constrtaints-marginal-2-2}
\sum_{l = n-1}^{j_1-2} p_{+l} \le \lambda_1 + \sum_{k = n}^{j_1-1} p_{k+}
\quad
(n=1, \ldots, j_1-1).
\end{equation}
The proof is thus complete because \eqref{eq:constrtaints-marginal-final-1}
holds by \eqref{eq:constrtaints-marginal-2} and \eqref{eq:constrtaints-marginal-2-2}.
\end{proof}

\smallskip
\begin{proof}[Proof of Lemma \ref{le:3-b}]

If $m_1 > 1,$ we first repeatedly utilize Lemma \ref{le:2-a} and obtain 
\begin{equation*}
\delta_{j_1, m_1}
= \delta_{j_1, n} + \sum_{s = n}^{m_1-1} (p_{+, j_1+s-1} - p_{j_1+s, +})
\quad
(n = 1, \ldots, m_1-1).
\end{equation*}
Because 
$
\delta_{j_1, m_1}
\le
\delta_{j_1, n}
$
for 
$
n=1, \ldots, m_1 - 1,
$
\begin{equation}
\label{eq:constrtaints-marginal-0}
\sum_{l=j_1+n-1}^{j_1+m_1-2} p_{+l} \le \sum_{k=j_1+n}^{j_1+m_1-1} p_{k+}
\quad
(n = 1, \ldots, m_1-1);
\end{equation}
Moreover, by repeatedly utilizing Lemma \ref{le:2-b}, we have 
\begin{equation*}
\delta_{j, m_1}
= \delta_{n, j_1+m_1-n} + \sum_{s = j_1}^{n-1} (p_{s+} - p_{+, s-1})
\quad
(n=j_1+1, \ldots, j_1+m_1-1)
\end{equation*}
Because 
$
\delta_{j_1, m_1}
\le 
\delta_{n, j_1 + m_1  - n}
$
for 
$
n=j_1+1, \ldots, j_1+m_1-1
$
\begin{equation*}
\sum_{k = j_1}^n p_{k+} \le \sum_{l = j_1-1}^{n-1} p_{+l}
\quad
(n=j_1, \ldots, j_1+m_1-2),
\end{equation*}
or equivalently, by the definition of $\lambda_1$ in \eqref{eq:define-lambda},
\begin{equation*}
\sum_{l = n}^{j_1+m_1-2} p_{+l} \le \sum_{k = n+1}^{j_1+m_1-1} p_{k+} - \lambda_1
\quad
(n=j_1, \ldots, j_1+m_1-2),
\end{equation*}
or equivalently
\begin{equation}
\label{eq:constrtaints-marginal-3}
\sum_{l = j_1 + n - 1}^{j_1+m_1-2} p_{+l} \le \sum_{k = j_1 + n}^{j_1+m_1-1} p_{k+} - \lambda_1
\quad
(n = 1, \ldots, m_1-1),
\end{equation}
The proof is thus complete because \eqref{eq:constrtaints-marginal-final-2} holds
by \eqref{eq:constrtaints-marginal-0} and \eqref{eq:constrtaints-marginal-3}.
\end{proof}

\smallskip
\begin{proof}[Proof of Lemma \ref{le:3-c}]

If $j_1+m_1 < J,$ we first repeatedly utilize Lemma \ref{le:2-a} and obtain 
\begin{equation*}
\delta_{j_1, m_1}
= \delta_{j_1, n} + \sum_{s = m_1}^{n-1} (p_{j_1+s, +} - p_{+, j_1+s-1})
\quad
(n=m_1+1, \ldots, J-j_1).
\end{equation*}
Because 
$
\delta_{j_1, m_1}
\le
\delta_{j_1, n}
$
for 
$
n = m_1+1, \ldots, J-j_1,
$
\begin{equation}
\label{eq:constrtaints-marginal-1}
\sum_{k=j_1+m_1}^{j_1+n-1} p_{k+} \le \sum_{l=j_1+m_1-1}^{j_1+n-2} p_{+l}
\quad
(n = m_1+1, \ldots, J-j_1).
\end{equation}
Moreover, by \eqref{eq:delta-iteration-1} for all $j_1=1, \ldots, J-1,$
\begin{equation}
\label{eq:delta-iteration-2}
\delta_{j_1, m_1} 
= \delta_{1, j_1+m_1-1} + p_{0+} + \sum_{k=j_1}^{J-1}p_{k+} - \sum_{l=j_1-1}^{J-1}p_{+l}.
\end{equation}
In addition, by Lemma \ref{le:2-c}
\begin{equation}
\label{eq:delta-iteration-3}
\delta_{1, j_1+m_1-1} = \delta_{j_1+m_1, J-j_1-m_1} + (p_{+, J-1} - p_{0+}).
\end{equation}
By combining \eqref{eq:delta-iteration-2} and \eqref{eq:delta-iteration-3}, and then repeatedly utilizing Lemma \ref{le:2-a}, we have 
\begin{align}
\label{eq:delta-iteration-5}
 {\delta _{{j_1},{m_1}}} &= {\delta _{{j_1} + {m_1},J - {j_1} - {m_1}}} + \sum\limits_{k = {j_1}}^{J - 1} {{p_{k + }}}  - \sum\limits_{l = {j_1} - 1}^{J - 2} {{p_{ + l}}} \nonumber \\ 
  &= {\delta _{{j_1} + {m_1},n}} + \sum\limits_{l = {j_1} + {m_1} - 1 + n}^{J - 2} {{p_{ + l}}}  - \sum\limits_{k = {j_1} + {m_1} + n}^{J - 1} {{p_{k + }}}  + \sum\limits_{k = {j_1}}^{J - 1} {{p_{k + }}}  - \sum\limits_{l = {j_1} - 1}^{J - 2} {{p_{ + l}}}  \nonumber \\ 
  &= {\delta _{{j_1} + {m_1},n}} + \sum\limits_{k = {j_1}}^{{j_1} + {m_1} - 1 + n} {{p_{k + }}}  - \sum\limits_{l = {j_1} - 1}^{{j_1} + {m_1} - 2 + n} {{p_{ + l}}} 
\quad
(n=1, \ldots, J-j_1-m_1) .
 \end{align}
Because 
$
\delta_{j_1, m_1}
\le 
\delta _{{j_1} + {m_1},n}
$
for 
$
n=1, \ldots, J-j_1-m_1,
$
\begin{equation*}
\sum_{k=j_1}^{j_1+m_1-1+n}p_{k+} \le \sum_{l=j_1-1}^{j_1+m_1-2+n}p_{+l}
\quad
(n=1, \ldots, J-j_1-m_1) . 
\end{equation*}
By the definition of $\lambda_1$ in \eqref{eq:define-lambda}, we can re-write the above inequalities as
\begin{equation*}
\sum_{k=j_1+m_1}^{j_1+m_1-1+n}p_{k+} \le \sum_{l=j_1+m_1-1}^{j_1+m_1-2+n}p_{+l} - \lambda_1
\quad
(n=1, \ldots, J-j_1-m_1), 
\end{equation*}
or equivalently
\begin{equation}
\label{eq:constrtaints-marginal-4}
\sum_{k=j_1+m_1}^{j_1+n-1}p_{k+} \le \sum_{l=j_1+m_1-1}^{j_1+n-2}p_{+l} - \lambda_1
\quad
(n = m_1 + 1, \ldots, J-j_1).
\end{equation}
The proof is thus complete because \eqref{eq:constrtaints-marginal-final-3}
holds by \eqref{eq:constrtaints-marginal-1} and \eqref{eq:constrtaints-marginal-4}.
\end{proof}

\section{Proof of Theorem \ref{thm:main}}\label{sec:theorem}

We prove Theorem \ref{thm:main} in two steps. First, we show $\gamma_U$ is indeed an upper bound. Second, we show the sharpness of $\gamma_U,$ by constructing a probability matrix $\bm{P}$ attaining it. As mentioned previously, in general there can be multiple probability matrices attaining $\gamma_U.$

\subsection{Step 1: Proving the upper bound}
For a fixed $j \in \{1, \ldots, J-1\},$
\begin{align*}
\mathrm{pr} \{ Y_i(1) > Y_i(0) \}
&= \mathrm{pr} \{ Y_i(1) \ge j, Y_i(0) < Y_i(1) \} + \mathrm{pr} \{ Y_i(1) < j, Y_i(0) < Y_i(1) \} \\
&= \mathrm{pr} \{ Y_i(1) \ge j, Y_i(0) \le j-1, Y_i(0) < Y_i(1) \} + \mathrm{pr} \{ Y_i(1),  Y_i(0) \ge j, Y_i(0) < Y_i(1) \} \\
&+ \mathrm{pr} \{ Y_i(1) < j, Y_i(0) < Y_i(1) \} \\
&= \underbrace{\mathrm{pr} \{ Y_i(1) \ge j, Y_i(0) \le j-1 \}}_{T_1} 
+ \underbrace{\mathrm{pr} \{ Y_i(1),  Y_i(0) \ge j, Y_i(0) < Y_i(1) \}}_{T_2} \\
&+ \underbrace{\mathrm{pr} \{ Y_i(1) < j, Y_i(0) < Y_i(1) \}}_{T_3}.
\end{align*}
By switching the labels of treatment and control, we obtain from the above identity that 
\begin{align*}
\mathrm{pr} \{ Y_i(1) < Y_i(0) \}
&= \underbrace{\mathrm{pr} \{ Y_i(0) \ge j, Y_i(1) \le j-1 \}}_{T_4} 
+ \underbrace{\mathrm{pr} \{ Y_i(0),  Y_i(1) \ge j, Y_i(1) < Y_i(0) \}}_{T_5} \\
&+ \underbrace{\mathrm{pr} \{ Y_i(0) < j, Y_i(1) < Y_i(0) \}}_{T_6}.   
\end{align*}
Therefore, by the definition of $\gamma$ in \eqref{eq:alpha}, 
\begin{align}
\label{eq:alpha-proof-0}
\gamma 
&= \mathrm{pr} \{ Y_i(1) > Y_i(0) \} - \mathrm{pr} \{ Y_i(1) < Y_i(0) \} \nonumber \\
&= (T_1 - T_4) + T_2 + T_3 - T_5 - T_6 \nonumber \\
&\le (T_1 - T_4) + T_2 + T_3. 
\end{align}
Below we deal with the three terms in \eqref{eq:alpha-proof-0}, namely $T_1 - T_4$, $T_2$ and $T_3$ separately. First, 
\begin{align}
\label{eq:alpha-proof-1}
T_1 - T_4
&= \mathrm{pr} \{ Y_i(1) \ge j, Y_i(0) \le j-1 \} - \mathrm{pr} \{ Y_i(0) \ge j, Y_i(1) \le j-1 \} \nonumber \\
&= \mathrm{pr} \{ Y_i(1) \ge j\} - \mathrm{pr} \{ Y_i(1) \ge j, Y_i(0) \ge j \} - \mathrm{pr} \{ Y_i(0) \ge j\} + \mathrm{pr} \{ Y_i(0) \ge j, Y_i(1) \ge j \} \nonumber \\
&= \Delta_j.
\end{align}
Second, for fixed $m \in \{1, \ldots, J-j\},$
\begin{align}
\label{eq:alpha-proof-2}
T_2
&= \mathrm{pr} \{ Y_i(1),  Y_i(0) \ge j, Y_i(0) < Y_i(1) \} \nonumber \\
&= \mathrm{pr} \{j \le Y_i(0) \le j+m-2, Y_i(0) < Y_i(1) \} + \mathrm{pr} \{j+m-1 \le Y_i(0) \le J-2, Y_i(0) < Y_i(1) \} \nonumber \\
&= \sum_{l=j}^{j+m-2} \sum_{k=l+1}^{J-1} p_{kl} 
+ \sum_{l=j+m-1}^{J-2} \sum_{k=l+1}^{J-1} p_{kl} \nonumber \\
&= \sum_{l=j}^{j+m-2} \sum_{k=l+1}^{J-1} p_{kl} 
+ \sum_{k=j+m}^{J-1} \sum_{l=j+m-1}^{k-1} p_{kl} \nonumber \\
&\le \sum_{l=j}^{j+m-2} p_{+l} + \sum_{k=j+m}^{J-1} p_{k+}.
\end{align}
Third,
\begin{align}
\label{eq:alpha-proof-3}
T_3
= \mathrm{pr} \{ Y_i(1) \le j-1, Y_i(0) < Y_i(1) \}  
= \sum_{k=1}^{j-1} \sum_{l=0}^{k-1} p_{kl} = \sum_{l=0}^{j-2} \sum_{k=l+1}^{j-1} p_{kl}  
\le \sum_{l=0}^{j-2} p_{+l}.
\end{align}
Therefore, by \eqref{eq:delta} and \eqref{eq:alpha-proof-0}--\eqref{eq:alpha-proof-3} we have proved that
$
\gamma \leq  \delta_{jm}.
$

\subsection{Step 2: Proving the sharpness}

This step consists of two parts. First, by the definition of $(j_1, m_1)$ in \eqref{eq:the-minimum-tuple} and Lemmas \ref{le:1}--\ref{le:3}, we construct a $J\times J$ matrix $\bm P = (p_{kl})_{0 \le k,l \le J-1}.$ Second, we prove that $\bm P$ is a well-defined probability matrix attaining the upper bound $\gamma_U,$ i.e., it has non-negative entries, that its row and column sums are $\bm p_1$ and $\bm p_0$ respectively, and that its corresponding relative treatment effect $\gamma$ is indeed $\delta_{j_1, m_1}.$

\subsubsection{Construction of the probability matrix}
For \emph{initialization}, we let $p_{kl}=0$ for all $k,l = 0, \ldots, J-1.$ Then, we use Lemma \ref{le:3} to \emph{update} certain entries of $\bm P,$ based on the values of $j_1$ and $m_1.$  
\begin{enumerate}[label = (\Roman*)]
\item \label{construction-1} If $j_1 > 1,$ by \eqref{eq:define-q-1} and \eqref{eq:constrtaints-marginal-final-1}, we apply Lemma \ref{le:1-a} to
$$
(q_{1+}, \ldots, q_{j_1-1, +})
\quad
\textrm{and}
\quad
(p_{+0}, \ldots, p_{+, j_1-2}),
$$
and \emph{update} the sub-matrix 
$
(p_{kl})_{1\le k \le j_1-1, 0 \le l \le j_1-2}
$ 
with non-negative entries such that it remains lower-triangular and satisfies
\begin{eqnarray}
\label{eq:submatrix-1-1}
\sum_{l^\prime=0}^{j_1-2} p_{kl^\prime} &\le & q_{k+}
\quad
(1\le k \le j_1-1), \\
\label{eq:submatrix-1-2}
\sum_{k^\prime=1}^{j_1-1} p_{k^\prime l} &=& p_{+l}
\quad
(0 \le l \le j_1-2). 
\end{eqnarray}

\item \label{construction-2} If $m_1 > 1,$ by \eqref{eq:define-q-2} and \eqref{eq:constrtaints-marginal-final-2}, we apply Lemma \ref{le:1-a} to
$$
(q_{j_1+1, +}, \ldots, q_{j_1+m_1-1, +})
\quad
\textrm{and}
\quad
(p_{+, j_1}, \ldots, p_{+, j_1+m_1-2}),
$$
and \emph{update} the sub-matrix 
$
(p_{kl})_{j_1+1 \le k \le j_1+m_1-1, j_1 \le l \le j_1+m_1-2}
$ 
with non-negative entries such that it remains lower-triangular and satisfies
\begin{eqnarray}
\label{eq:submatrix-2-1}
\sum_{l^\prime=j_1}^{j_1+m_1-2} p_{kl^\prime} & \le & q_{k+}
\quad
(j_1+1 \le k \le j_1+m_1-1), \\ 
\label{eq:submatrix-2-2}
\sum_{k^\prime=j_1+1}^{j_1+m_1-1} p_{k^\prime l} &=& p_{+l}
\quad
(j_1 \le l \le j_1+m_1-2). 
\end{eqnarray}

\item \label{construction-3} If $j_1 + m_1 < J,$ by \eqref{eq:define-q-3} and \eqref{eq:constrtaints-marginal-final-3}, we apply Lemma \ref{le:1-c} to
$$
(p_{j_1+m_1, +}, \ldots, p_{J-1, +})
\quad
\textrm{and}
\quad
(q_{+, j_1+m_1-1}, \ldots, q_{+, J-2}),
$$
and \emph{update} the sub-matrix 
$
(p_{kl})_{j_1+m_1 \le k \le J-1, j_1+m_1-1 \le l \le J-2}
$ 
with non-negative entries such that it remains lower-triangular and satisfies
\begin{eqnarray}
\label{eq:submatrix-3-1}
\sum_{l^\prime=j_1+m_1-1}^{J-2} p_{kl^\prime} &=& p_{k+},
\quad
(j_1+m_1 \le k \le J-1), \\
\label{eq:submatrix-3-2}
\sum_{k^\prime=j_1+m_1}^{J-1} p_{k^\prime l}  &\le & q_{+l}
\quad
(j_1+m_1-1 \le l \le J-2).
\end{eqnarray}
\end{enumerate}

We further \emph{update} $\bm P$ in the following sequential fashion. 
\begin{enumerate}[label = (\Roman*)]
\setcounter{enumi}{3}
\item \label{construction-4} Let 
\begin{equation}
\label{eq:fill-in-1}
p_{j_1+m_1-1, j_1+m_1-1} = \max(0, \lambda_1). 
\end{equation}

\item \label{construction-5} For each $k=j_1, \ldots, j_1+m_1-1,$ let
\begin{equation}
\label{eq:fill-in-2}
p_{k, j_1-1} = p_{k+} - \sum_{l^\prime=j_1}^{J-1}p_{kl^\prime}. 
\end{equation}

\item \label{construction-6} Let 
\begin{equation}
\label{eq:fill-in-3}
p_{j_1-1, j_1-1} = p_{+, j_1-1} - \sum_{k^\prime = j_1}^{j_1+m_1-1} p_{k^\prime, j_1-1}. 
\end{equation}

\item \label{construction-7} For all $k=0, \ldots, j_1-1$ and $l = j_1+m_1-1, \ldots, J-1,$ let
\begin{equation}
\label{eq:fill-in-leftover}
p_{kl} = 
\left(
p_{k+} - \sum_{l^\prime = 0}^{j_1+m_1-2} p_{kl^\prime}
\right)
\left(
p_{+l} - \sum_{k^\prime = j_1}^{J-1} p_{k^\prime l}
\right).
\end{equation}

\end{enumerate}

\begin{figure}[ht]
    \centering
	\includegraphics[width = 1.1\textwidth, left]{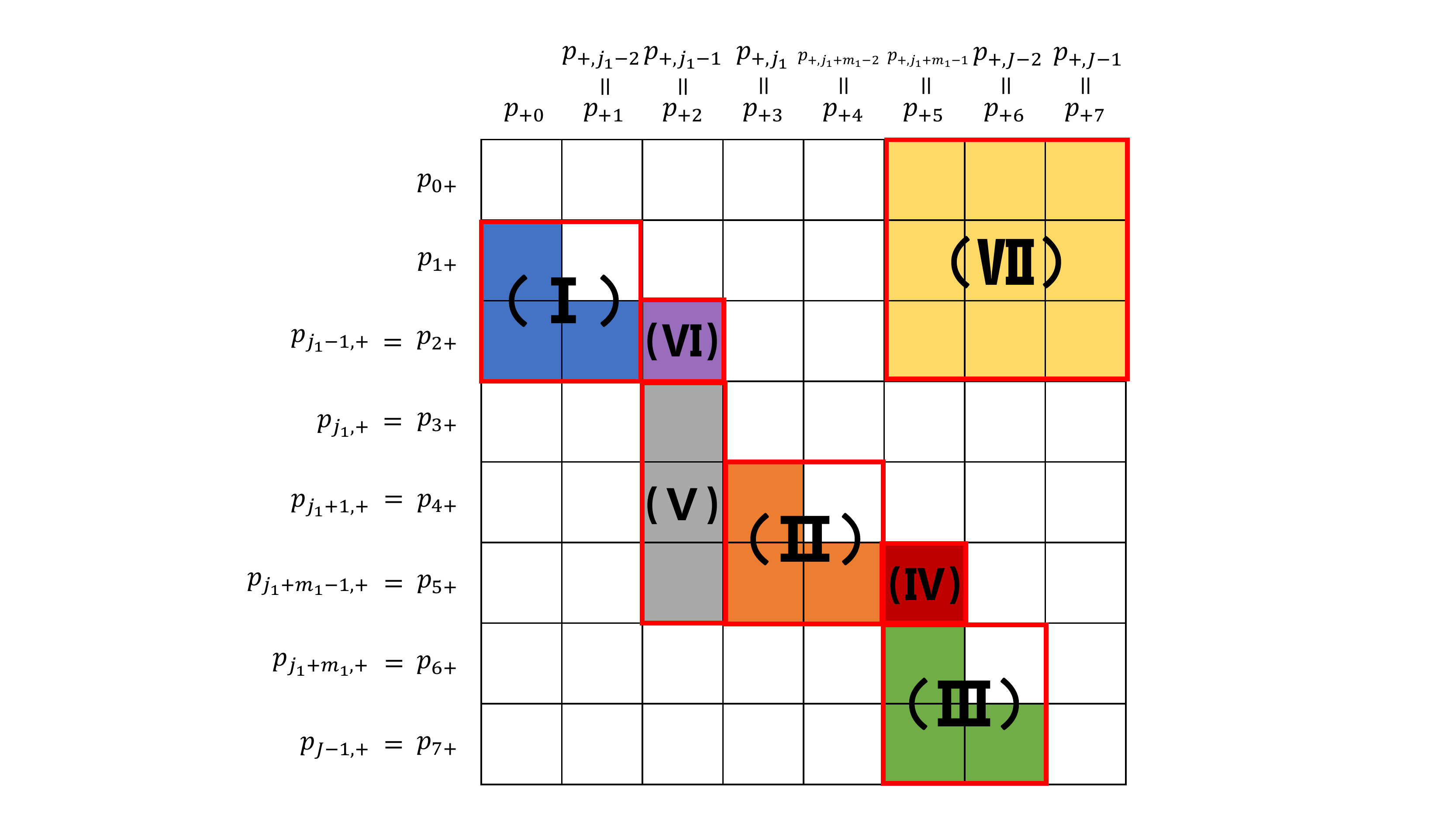}	
	\caption{Visualization of the construction of $\bm P,$ when $J=8,$ $j_1 = 3$ and $m_1 = 2.$ The uncolored entries are all zeros. The colored entries are defined by steps \ref{construction-1}---\ref{construction-7}, correspondingly. }
	\label{fig:construction}
\end{figure}

To summarize, our construction procedure is defined by steps \ref{construction-1}---\ref{construction-7}; to be more specific, equations \eqref{eq:submatrix-1-1}--\eqref{eq:fill-in-leftover}. Figure \ref{fig:construction} contains a visual illustration of the construction of the probability matrix, where $J=8,$ $j_1 = 3$ and $m_1 = 2.$

\subsubsection{Validation of the probability matrix}  

\paragraph{Non-negative entries} We verify that all entries of the probability matrix $\bm P,$ defined by steps \ref{construction-1}--\ref{construction-7}, are non-negative. 

\begin{enumerate}
\item All entries defined in steps \ref{construction-1}---\ref{construction-4} are non-negative by definition. 

\item For entries defined in step \ref{construction-5}, i.e., $p_{k, j_1 - 1}$ for all $k=j_1, \ldots, j_1 + m_1 - 1,$ we discuss two cases. First, if $m_1 = 1,$ by \eqref{eq:define-lambda} and \eqref{eq:fill-in-1} we have
$
p_{j_1, j_1} 
= 
\max
\left(
0, p_{j_1, +} - p_{+, j_1-1}
\right)
\le p_{j_1, +},
$
which implies that $p_{j_1, j_1 - 1} \ge 0.$ Second, if $m_1 > 1,$ by \eqref{eq:submatrix-2-1} and definitions of the $q_{k+}$'s in \eqref{eq:define-q-2},
$
p_{k, j_1 - 1} \ge 0
$
for all
$
k = j_1, \ldots, j_1 + m_1 - 2.
$
Therefore, we only need to prove that 
\begin{equation*}
p_{j_1+m_1-1, j_1-1} = p_{j_1+m_1-1, +} - \sum_{l=j_1}^{j_1+m_1-1} p_{j_1+m_1-1, l} \ge 0.
\end{equation*}
This is guaranteed by \eqref{eq:define-q-2} and \eqref{eq:submatrix-2-1}, because
\begin{equation*}
\sum_{l=j_1}^{j_1+m_1-2}p_{j_1+m_1-1, l} + p_{j_1+m_1-1, j_1+m_1-1} 
\le q_{j_1+m_1-1, +} + \max(0, \lambda_1)
= p_{j_1+m_1-1, +}.
\end{equation*}

\item For $p_{j_1-1, j_1-1}$ defined in step \ref{construction-6}, by \eqref{eq:define-lambda}, \eqref{eq:submatrix-2-2}, \eqref{eq:fill-in-2} and \eqref{eq:fill-in-3},
\begin{align}
\label{eq:define-p-j-1-j-1}
p_{j_1-1, j_1-1} 
&= p_{+, j_1-1} - \sum_{k = j_1}^{j_1+m_1-1} 
\left(
 p_{k+} - \sum_{l^\prime=j_1}^{J-1}p_{kl^\prime}
\right) \nonumber \\
&= p_{+, j_1-1} - \sum_{k = j_1}^{j_1+m_1-1} p_{k+} + \sum_{l=j_1}^{j_1+m_1-2}p_{+l} + p_{j_1+m_1-1, j_1+m_1-1} \nonumber \\
&= \sum_{l=j_1-1}^{j_1+m_1-2}p_{+l} - \sum_{k = j_1}^{j_1+m_1-1} p_{k+} + \max(0, \lambda_1) \nonumber \\
&= -\lambda_1 + \max(0, \lambda_1) \nonumber \\
&= \max(0, -\lambda_1) \ge 0;
\end{align}
 
\item To prove all entries defined by step \ref{construction-7} are non-negative, we will prove that 
\begin{eqnarray}
\label{eq:leftover-guarantee-1}
p_{k+} - \sum_{l^\prime = 0}^{j_1+m_1-2} p_{kl^\prime} &\ge & 0
\quad
(k=0, \ldots, j_1-1),\\
\label{eq:leftover-guarantee-2}
p_{+l} - \sum_{k^\prime = j_1}^{J-1} p_{k^\prime l} &\ge & 0
\quad
(l = j_1+m_1-1, \ldots, J-1).
\end{eqnarray}

\begin{enumerate}

\item First, we prove \eqref{eq:leftover-guarantee-1}.
By the fact that
$$
p_{0l} = 0
\quad
(l = 0, \ldots, j_1 + m_1 - 2),
$$
the definitions of 
$
q_{1+}, \ldots, q_{j_1-1, +}
$ 
in \eqref{eq:define-q-1}, and \eqref{eq:submatrix-1-1}, it is straightforward to verify that \eqref{eq:leftover-guarantee-1} holds for all
$
k \in \{0, \ldots, j_1 - 1\} \backslash \{j_1 - 1\}.
$ 
To further prove that
$$
p_{j_1-1, +} - \sum_{l=0}^{j_1-1}p_{j_1-1, l} \ge 0,
$$
we discuss two cases:
\begin{enumerate}
\item If $j_1 > 1,$ by \eqref{eq:define-q-1}, \eqref{eq:submatrix-1-1}, and \eqref{eq:define-p-j-1-j-1},
\begin{equation*}
\sum_{l=0}^{j_1-2} p_{j_1-1, l} + p_{j_1-1, j_1-1} 
\le q_{j_1-1, +} + \max(0, -\lambda_1)
= p_{j_1-1, +};
\end{equation*}

\item If $j_1 = 1,$ by  \eqref{eq:define-lambda} and \eqref{eq:define-p-j-1-j-1} we only need to prove
\begin{equation*}
p_{0+} + \underbrace{\sum_{k=1}^{m_1} p_{k+} - \sum_{l=0}^{m_1-1} p_{+l}}_{\lambda_1} \ge 0.
\end{equation*}
If $m_1 = J-1,$ the left side
\begin{align*}
{p_{0 + }} + {\lambda _1} = {p_{0 + }} + \sum\limits_{k = 1}^{J - 1} {{p_{k + }}}  - \sum\limits_{l = 0}^{J - 2} {{p_{ + l}}} 
 = 1 - \sum\limits_{l = 0}^{J - 2} {{p_{ + l}}} 
 = {p_{ + ,J - 1}} \ge 0
\end{align*}
If $m_1 < J-1,$ its equivalent form
$
\sum_{k=m_1+1}^{J-1}p_{k+} \le \sum_{l=m_1}^{J-1} p_{+l}
$
holds by \eqref{eq:constrtaints-marginal-1}.
\end{enumerate}

\item Second, we prove \eqref{eq:leftover-guarantee-2}.
By the fact that
$$
p_{k, J-1} = 0
\quad
(k = 0, \ldots, j_1 + m_1 - 2),
$$
the definitions of 
$
q_{+, j_1 + m_1 - 1}, \ldots, q_{+, J - 2}
$ 
in \eqref{eq:define-q-3}, and \eqref{eq:submatrix-3-2}, it is straightforward to verify that \eqref{eq:leftover-guarantee-2} holds for 
$
l \in \{j_1 + m_1 - 1, \ldots, J - 1\} \backslash \{j_1 + m_1 - 1\}.
$ 
To further prove that
$$
p_{+, j_1+m_1-1} - \sum_{k=j_1+m_1-1}^{J-1} p_{k, j_1+m_1-1} \ge 0,
$$
we discuss two cases:
\begin{enumerate}
\item If $j_1+m_1 < J,$ by \eqref{eq:define-q-3} and \eqref{eq:submatrix-3-2},
\begin{equation*}
\sum_{k=j_1+m_1}^{J-1} p_{k, j_1+m_1-1} + p_{j_1+m_1-1, j_1+m_1-1} 
\le q_{+, j_1+m_1-1} + \max(0, \lambda_1)
= p_{+, j_1+m_1-1};
\end{equation*}

\item If $j_1+m_1 = J,$ by  \eqref{eq:define-lambda} and \eqref{eq:fill-in-1}, we only need to prove that
\begin{equation*}
\underbrace{\sum_{k=j_1}^{J-1} p_{k+} - \sum_{l=j_1-1}^{J-2} p_{+l}}_{\lambda_1}
\le
p_{+, J-1}.
\end{equation*}
If $j_1 = 1,$ note that the left side
\begin{align*}
{\lambda _1} = \sum\limits_{k = 1}^{J - 1} {{p_{k + }}}  - \sum\limits_{l = 0}^{J - 2} {{p_{ + l}}} 
 = {p_{ + ,J - 1}} - {p_{0 + }} \le {p_{ + ,J - 1}}
\end{align*}

If $j_1 > 1,$ its equivalent form
$
\sum_{l=0}^{j_1-2}p_{+l} \le \sum_{k=0}^{j_1-1} p_{k+}
$
holds by \eqref{eq:constrtaints-marginal-2}.
\end{enumerate}

\end{enumerate}

\end{enumerate}

\paragraph{Correct row and column sums} To verify the column and row sums, note that by \eqref{eq:submatrix-1-2}, \eqref{eq:submatrix-2-2}, \eqref{eq:fill-in-3} and \eqref{eq:fill-in-leftover}, the column sums of $\bm P$ are $p_{+0}, \ldots, p_{+, J-1},$ respectively. Similarly, by \eqref{eq:submatrix-3-1}, \eqref{eq:fill-in-2} and \eqref{eq:fill-in-leftover}, the row sums of $\bm P$ are $p_{0+}, \ldots, p_{J-1, +},$ respectively.

\paragraph{The relative treatment effect $\gamma$ of the constructed $\bm{P}$ attains the upper bound $\gamma_U$} To prove that the relative treatment effect of $\bm P$ is indeed $\delta_{j_1, m_1},$ note that $\bm P$ is initialized by all zeros, and that the sub-matrices constructed in steps \ref{construction-1}---\ref{construction-3} are all lower-triangular, which means that:
\begin{align*}
\mathop{\sum\sum}_{ k > l}p_{kl} 
&=
\underbrace{ \sum_{k=1}^{j_1-1}\sum_{l=0}^{k-1}p_{kl}}_{\textrm{(I)}}
+ \underbrace{\sum_{k=j_1}^{j_1+m_1-1}\sum_{l=j_1-1}^{k-1}p_{kl}}_{\textrm{(II)}}
+  \underbrace{ \sum_{k=j_1+m_1}^{J-1}\sum_{l=j_1+m_1-1}^{k-1}p_{kl}}_{\textrm{(III)}}, \\
\mathop{\sum\sum}_{ k < l}p_{kl}  
&=
\underbrace{\sum_{k=0}^{j_1-1}\sum_{l=j_1+m_1-1}^{J-1}p_{kl}}_{\textrm{(IV)}}.
\end{align*}
By \eqref{eq:submatrix-1-2} and \eqref{eq:submatrix-3-1}
\begin{equation}
\label{eq:calculate-alpha-I-and-III}
\textrm{(I)} 
= \sum_{l=0}^{j_1-2}p_{+l},
\quad
\textrm{(III)} 
= \sum_{k=j_1+m_1}^{J-1} p_{k+}.
\end{equation}
By \eqref{eq:fill-in-2} and the initialization with zeros,
\begin{equation}
\label{eq:calculate-alpha-II}
\textrm{(II)} 
= \sum_{k=j_1}^{j_1+m_1-1}p_{k+} - p_{j_1+m_1-1, j_1+m_1-1}.
\end{equation}
By \eqref{eq:submatrix-2-2} and \eqref{eq:fill-in-2}--\eqref{eq:fill-in-leftover},
\begin{align}
\label{eq:calculate-alpha-IV}
\textrm{(IV)}
&= \sum_{k=0}^{j_1-1}p_{k+}
- \sum_{k=1}^{j_1-1}\sum_{l=0}^{k-1}p_{kl} - p_{j_1-1, j_1-1} \nonumber \\
&=
\sum_{k=0}^{j_1-1}p_{k+}
-
\sum_{l=0}^{j_1-2}p_{+l}
-
\left(
p_{+, j_1-1} - \sum_{k = j_1}^{j_1+m_1-1} p_{k, j_1-1}
\right) \nonumber \\
&= \sum_{k = j_1}^{j_1+m_1-1} p_{k, j_1-1} + \sum_{k=0}^{j_1-1}p_{k+} - \sum_{l=0}^{j_1-1}p_{+l} \nonumber \\
&= \sum_{k=j_1}^{j_1+m_1-1}p_{k+} -  \sum\limits_{k = {j_1 + 1}}^{{j_1} + {m_1} - 1} {\sum\limits_{l = {j_1}}^{{j_1} + {m_1} - 2} {{p_{kl}}} } - p_{j_1+m_1-1, j_1+m_1-1} - \Delta_{j_1}. \nonumber \\
&= \sum_{k=j_1}^{j_1+m_1-1}p_{k+} - \sum\limits_{l = {j_1}}^{{j_1} + {m_1} - 2} {\sum\limits_{k = {j_1 + 1}}^{{j_1} + {m_1} - 1} {{p_{kl}}} } - p_{j_1+m_1-1, j_1+m_1-1} - \Delta_{j_1}. \nonumber \\
&= \sum_{k=j_1}^{j_1+m_1-1}p_{k+} - \sum_{l=j_1}^{j_1+m_1-2}p_{+l} - p_{j_1+m_1-1, j_1+m_1-1} - \Delta_{j_1}.
\end{align}
Consequently, by \eqref{eq:calculate-alpha-I-and-III}--\eqref{eq:calculate-alpha-IV},
\begin{align*}
\gamma 
= \textrm{(I)} + \textrm{(II)} + \textrm{(III)} - \textrm{(IV)} 
 = \Delta_{j_1} + \sum_{l=0}^{j_1-2}p_{+l} +  \sum_{k=j_1+m_1}^{J-1} p_{k+} 
+ \sum_{l=j_1}^{j_1+m_1-2} p_{+l} 
 = \delta_{j_1, m_1}.
\end{align*}

\section{Proof of Corollary \ref{coro:sharp-lower-bound}}\label{sec:corollary}

By switching the labels of the treatment and control, the relative treatment effect becomes 
$
\gamma^\prime 
= 
\textrm{pr}\left\{ {{Y_i}\left( 0 \right) > {Y_i}\left( 1 \right)} \right\} 
- 
\textrm{pr} \left\{ {{Y_i}\left( 0 \right) < {Y_i}\left( 1 \right)} \right\} 
=  - \gamma. 
$
Because 
$
\Delta_j^\prime 
= 
\textrm{pr} \left\{ {Y_i}\left( 0 \right) \ge j \right\} 
- 
\textrm{pr} \left\{ {Y_i}\left( 1 \right) \ge j \right\} 
=  - {\Delta_j},
$ 
using \eqref{eq:xi} we obtain 
\begin{equation*}
\delta_{jm}^\prime
=  
- \Delta_j 
+ \sum\limits_{k = 0}^{j - 2} p_{k + }  
+ \sum\limits_{l = j + m}^{J - 1} p_{ + l}  
+ \sum\limits_{k = j}^{j + m - 2} p_{k + }  
=  
- \xi_{jm}.
\end{equation*}
Using Theorem \ref{thm:main}, we obtain that the sharp upper bound on 
$\gamma '=  - \gamma$
is
$
\min_{1 \le j \le J - 1} ~ \min_{1 \le m \le J - j} (- \xi_{jm}),
$
which completes the proof.

\end{document}